\documentclass[showpacs,floatfix]{revtex4}
\usepackage[dvips]{graphicx}
\usepackage{epsfig}

\usepackage[cp1251]{inputenc}
\usepackage[english, russian]{babel}
\usepackage{color}
\usepackage{ulem}
\usepackage{bm}

\usepackage{feyn}

\begin{document}

\title{Simple analysis of scattering data with Ornstein-Zernike equation}

\author{E.I.Kats$^{1}$, A.R.Muratov$^{2,3}$}

\affiliation{$^1$ Landau Institute for Theoretical Physics, RAS, \\
142432, Chernogolovka, Moscow region, Russia \\
$^2$ Oil and Gas Research Institute, RAS,  \\
Gubkina st. 3, Moscow, Russia, 119333\\
$^3$ Gubkin State University of Oil and Gas,\\
Leninsky Prospekt, 65, Moscow B-296, GSP-1, 119991, Russia}

\begin{abstract}

In this paper we propose and explore a method of analysis of the scattering experimental data for uniform liquid-like
systems. In our pragmatic approach we are not trying to introduce by hands an artificial small parameter
to work out a perturbation theory with respect to the known results e.g., for hard spheres or sticky-hard  spheres (all the more that in the agreement with the notorious Landau statement, there is no any physical  small parameter for liquids). Instead of it guided by the experimental data we are solving the the Ornstein-Zernike equation with a trial (variational) form of the inter-particle interaction potential. To find all needed correlation functions this variational input is iterated numerically to satisfy the Ornstein-Zernike equation supplemented by a closure relation. We illustrate by a number of model and real experimental examples of the X-ray and neutron  scattering data how the approach works.

\end{abstract}


\maketitle

\section{Introduction}
\label{int}

Liquids (and not only so-called complex liquids or colloidal suspensions  but as well not exotic at all conventional simple organic
and inorganic ones)  are not structureless at any scale uniform media. The knowledge of their structure is very essential for 
understanding of the underlying physics and chemistry, and it allows a non-blind search of new ''smart'' materials possessing properties required for various applications. Nowadays the topic is a
multidisciplinary area including many basic science problems
involving physics, chemistry, biology and
applications. Recently there has been an essential evolution in
our understanding of the structures and phase transitions in liquids. It can be illustrated merely by
continuously growing number of exciting new publications (some of
those will be cited in what follows in our paper). The progress, as always, is driven not only by developments of new
experimental techniques but also theoretical advances, promising potential applications, and related
interesting fundamental scientific problems. 

Experimentally complete and detailed structural information is obtained by a number of 
scattering methods (X-rays, neutrons, light). Results are summarized in many reviews and monographs, e.g., in multiple editions of the well known J.P. Hansen and I. R. McDonald book \cite{HM06}, containing also numerous relevant references.
However liquids as any studied in physics objects have so-to-speak two faces. First, mentioned above - experimental data,
and second - their theory description. Modern powerful computers and software enable to perform large scale simulations of molecular
liquids or colloidal dispersions. Thus, there are high precision scattering data and high accuracy simulations, and therefore it is tempting to think that nothing else is needed in the field. Unfortunately it is not completely true. The matter is that the full set of parameters
which determine the experimentally measured scattering intensity $I$ (or related to $I$ the static structure factor $S(q)$), and the parameters needed to perform numeric simulations are not exactly the same, and what is worse only barely known. Readers can find a lot of original publications, reviews and monographs from the theory and simulation side (just to mention a few, see
\cite{FS01}, \cite{HY08}, \cite{GC13}). One of the main difficulties in comparing the results of the large-scale simulations with
specific experimental measurements is the availability of an accurate connection between experimental control parameters and the theoretical variables needed for the simulations. The actual values of the parameters
are determined by the microscopic interactions, which are not well known. It might be not so important because in simulations the level of details is much greater then can be obtained experimentally, but there is another disadvantage of large scale simulations, they do not tell us which elements of the interactions are the most essential for a given system behavior. 
It should be cause for general embarrassment in the field
that there are no still answers on even the most basic questions on structural and thermodynamic properties of liquids.

To overcome somehow this mutual uncertainty of the scattering data and simulations, and to relate the data
with physical system characteristics we need the theory guideline. And here we face another trouble.
Theory in the rigoristic meaning of the word (see e.g., \cite{LL80}) may not be developed
for molecular or colloidal liquids, since there is no any small parameter. 
To find a way from the impasse we propose a pragmatic approach. The rigoristic theoretical view 
is certainly correct, however the heuristic theory approach combined with experimental input and common physical wisdom,
provides useful tool to describe experimental data with a few phenomenological parameters.
Moreover obtained theoretically results, can be also used for certain new predictions. We introduce some effective (model) inter-particle potentials, which we
regard as experimentally determined ones. Along this way to test such combined (theory-experiment) approach we need a standard reference system. For
the colloidal dispersions such reference system is the dispersion of hard spheres.
The fact is that structural and even dynamic properties of liquids are dominated by the molecular repulsive cores. This general deceptively simple Van der Waals observation leads to the idea that hard spheres is a suitable basic model of the liquid state.

The liquid structure factor which can be determined from the measured scattering
intensity is a Fourier transform of the pair density
correlation function, see details in the next section. The density correlation function satisfies the formally
exact Ornstein-Zernike (OZ) equation. Unfortunately the equation is not in a closed form, because it
contains two unknown functions, and to solve this equation one has to add a closure relation.
It turns out that the structure factor of the hard-spheres liquid can be accurately calculated using the  Percus-Yevick (PY) closure relation. Unfortunately the remainder (with respect to hard spheres) interactions may not always be treated as a perturbation. Then one has to rely on  different methods (see, e.g., \cite{GM01},
\cite{BW05}, \cite{CM05}, \cite{ZP14}). In this work we are solving the OZ equation for a smooth combination of the correlation functions (see details in the next section) and introduce an effective inter-particle interaction potential. To find the other correlation functions this input is iterated numerically to satisfy the OZ equation supplied with a closure relation \cite{HM06}. We test the approach investigating a number of model and real experimental examples of the X-ray and neutron  scattering data. 

Common experience in data fitting claims that for short-range interactions between the particles,
the  PY closure relation gives very reasonable description of the data. For more long-ranged interactions
the better description of the data is obtained by the so-called hypernetted-chain (HNC) closure relation \cite{HM06}.
Variation of external conditions and material parameters results in the change of physical
properties of the system. The static structure factor $S(q)$ is the main quantity one needs to analyze experimental data and to confront the data with the theory.
The OZ equation with PY closure
can be solved analytically only for hard spheres \cite{Wertheim}, \cite{Thiele}
or hard-spheres with a very short-range attraction (sticky hard sphere model)\cite{Baxter2},
and more recent experimental and theoretical advances and improvements can be found in \cite{LP11}, \cite{GV14}.
The first example (hard spheres) is very important but oversimplified to describe the real molecular liquids and colloidal dispersions.
As it concerns to the second model (sticky hard spheres) its application is strongly
limited \cite{Menon}. 
In fact it is a general drawback of the standard approach. The interpretation of the scattering data and the obtained
values of the parameters are model dependent, and heavily rely on the assumptions used in the data analysis.
Although the OZ integral equation can be solved by iterations \cite{HM06},
the method of the solution which provides the convergence and stability for
a general case is not proposed (see the documentation to SASFIT software \cite{SASFIT}). Stability of the algorithm can be obtained if one can guess in advance
the form of the solution with several adjustable parameters, and it is not a trivial task.

Thus, the desire to
understand the physical system characteristics behind its structure factor which is the main
aim of this paper, is hardly surprising.
The plan of our paper is as follows. In the next section \ref{theory} we describe the
main steps of our approach. We analyze the scattering data in the framework of the OZ equation.
Our method can be applied both for molecular liquids and for colloidal dispersions, but for the latter systems
it works if polydispersity is not too high. Then in section \ref{exp} we illustrate how our approach works and present
a number of model (with known form for the structure factor) and experimental results which we analyze by our method. Finally, in the section \ref{con} we summarize the main steps of our approach and the results of the work.

\section{Theory}
\label{theory}

Consider suspension of monodisperse hard spheres of diameter $\sigma$. Volume fraction occupied
by the spheres is $\phi = \pi \sigma^3{\bar n}/6$, where ${\bar n}$ is its average concentration.
The pair correlation function is defined as
$g(r)=1+{\bar n}h(r)=\langle n(r) n(0)\rangle /{\bar n}^2-{\bar n}\delta(r)$, and two other correlation functions are useful to describe the scattering data: the total correlation function $h(r)$ and the direct correlation function
$c(r)=- [{\delta^2 {\cal F}}/({T\delta n(r)\delta n(0)})]$. The functions $h(r)$ and $c(r)$ enter to the exact
OZ equation.
\begin{equation}
h({\bf x})=c({\bf x})+{\bar n}\int d^3y\, h({\bf y}) c({\bf x-y})\ .
\label{1}
\end{equation}
Static structure factor $S(q)$,  is determined as
\begin{equation}
S(q)=1+\frac{4\pi {\bar n}}q \int_0^{\infty} dr\,r sin(qr)h(r) \ .
\label{2}
\end{equation}
OZ equation can be rewritten in the Fourier representation as
\begin{equation}
S(q)=1+ {\bar n}h(q)=1/(1-{\bar n}c(q)) \ ,
\label{3}
\end{equation}
and to solve this equation it is necessary to add a closure relation.
Two the most popular (and turned out physically justified) closure relations are the PY closure equation
\begin{equation}
c(r)=(1-e^{\beta V(r)} )g(r)\ ,
\label{4}
\end{equation}
(where $V(r)$ is an interaction potential between particles and $\beta=(k_B T)^{-1}$), and
HNC closure looks like
\begin{equation}
g(r)=e^{\gamma (r)-\beta V(r)} \ ,
\label{5}
\end{equation}
where $\gamma(r)=h(r)-c(r)$.

If the form of the inter-particle potential $V(r)$ is known, the closure relation allows to solve the OZ equation
and then to compute $S(q)$. Unfortunately it is almost never the case, and one has either to guess about $V(r)$,
or try to find some insight by fitting the scattering data. It looks as a vicious circle (because to fit the data we
need to solve the OZ and the closure equations, what is impossible without the knowledge of the potential). Luckily the situation is not so hopeless, and the both tasks (namely to compute $S(q)$ and to guess on the form of $V(r)$) can be done
simultaneously by iterations. We chose first (guided by physical arguments) a model potential, then compute $S(q)$,
compare the results to the experimental data and repeat the procedure, until the agreement with experimental data becomes
satisfactory. What is important for our approach, that it is sufficient to know experimental data in a relatively narrow (around the
first peak in $I(q)$) range of the scattering wave vectors to compute the static structure factor in a much broader range of $q$. 

Usually it is supposed that the interaction
potential is very large and repulsive for small inter-particle distance $r<\sigma$, (where $\sigma $ stands for an
effective size of the particle hard core) $V(r)={\bar V}\gg k_B T$, and vanishes outside the interaction region
$R_{int}<r$, $V(r)=0$. It is easy to see that the PY and HNC closure relations imply  that
\begin{eqnarray}
&&g(r)\propto e^{-\beta \bar V}\ , \quad {\rm if} \ r<\sigma
\nonumber \\
&&c(r)=0\ , \quad {\rm if} \ R_{int} < r \ .
\label{6}
\end{eqnarray}
However, what is directly measured in any scattering experiment is not the static structure factor. The measured quantity is
the scattering intensity $I({\bf q}, {\bar n})$ (where as before ${\bf q}$ is the scattering wave vector, 
and ${\bar n}$ is the average particle concentration). For a very dilute dispersion, when 
${\bar n}={\bar n}_{dil}$ is small ${\bar n}_{dil}\sigma ^3 \ll 1$,
$I(q , {\bar n}_{dil})$ is the scattering intensity from a single particle, termed traditionally as the particle form factor. For
molecular liquids, or for colloidal dispersions with relatively small polydispersity (the narrow particle size distribution function), the static structure factor can be determined as
$S(q)=\bar n_{dil}\,I(q,\bar n)/(\bar n\,I(q,\bar n_{dil})) $

Unfortunately it is impossible to calculate accurately the correlation functions $h(r),\ c(r)$ 
in the $r$-space by the inverse Fourier transformation of the static structure factor, because $S(q)$ decreases too slow, 
usually as $q^{-1}$. Luckily, for the  function $\gamma (r)$ the situation is much better. It can be obtained by the Fourier
transformation. The reason is that $\gamma(r)$ is a smooth function unlike the total and direct correlation functions,
and its Fourier transform decreases fast, e.g., for the hard spheres like $1/q^3$.
The required range of the wave vectors in the experimentally measured scattering intensity depends on the interaction potential. For the hard
spheres at not too high particle volume fraction (say, $\phi \leq 0.4$)), it is sufficient to know the scattering intensity for $q < 10/\sigma$.
Then the function $\gamma(r)$ can be calculated directly from the static structure factor and the exact OZ equation
(without an explicit use of any closure equation):
\begin{equation}
\gamma(r)=\frac 1{2\pi^2 r\bar n}\int_0^{\infty}dq\, q \sin(qr)(S(q)-2+1/S(q)) \ .
\label{7}
\end{equation}
Eq. (\ref{7}) is the main message of our work. Of course to compute $c(r)$, $g(r)$, and $h(r)$ separately one has to supplement the OZ equation by one or another closure relation.

The equation for the direct correlation function for the PY closure reads as
\begin{equation}
c(r_i)=(1+\gamma(r_i))(e^{-\beta V(r_i)}-1)\ ,
\label{8}
\end{equation}
and for for the HNC closure it is
\begin{equation}
c(r_i)=e^{\gamma (r_i)-\beta V(r_i)}-1-\gamma(r_i) \ .
\label{9}
\end{equation}
Technically to solve the OZ equation we design a simple numerical method, similar to that proposed by
Gillan \cite{Gillan}. The original \cite{Gillan} method is based on the discrete form the of the equations and Fourier
transform of the correlation functions (instead of direct calculation of the integral in (\ref{1})).
By our code we are solving the following discretized equations
\begin{eqnarray}
&& \tilde c_j=\frac{4 \pi h_r}{q_j}\sum_{i=1}^N c_i r_i\sin(q_j r_i) \ ,
\nonumber \\
&& S_i'=\frac {h_q}{2\pi^2 r_i}\sum_{j=1}^N \frac {q_j\sin(q_j r_i)}{1-\bar n \tilde c_j} \ .
\label{10}
\end{eqnarray}
where $r_i=i\,h_r,\ q_j=j\,h_q,\ h_q=\pi/(N h_r)$, $h_r$ and $h_q$ are steps in $r$ and $q$ 
space correspondingly. Fitting the results obtained from the equations (\ref{7}-\ref{10}) to the
experimentally found the static structure factor, we are able also to determine the interaction potential,
calculate the density correlation function $g(r)$, and its main characteristic features, and compute as well
some thermodynamic properties of the system, e.g., the pressure. It is worth to cite here the work
\cite{WK10} where the inverse problem (to find the interaction potential from the scattering data)
has been also discussed.

\section{Analysis of experiment}
\label{exp}

As the first test of our approach we treat the hard sphere model data obtained by the exact solution of the
OZ and PY equations. This is our input ''experimental'' data. The test will allow also to estimate
the accuracy of our computation due to the limited range of the available wave vectors, and finite precision
of the discretized Fourier transform. In the test (in dimensionless $r$ measured in the units of $\sigma $)
we take the hard sphere volume fraction $\phi = 0.17$ and 200 points from the data-set of the exact solution 
for the structure factor $S(r)$ in the range of dimensionless $r$ $r/\sigma =1 \div 11 $.
To analyze these ''experimental'' data by our method, we should first find the effective interaction potential.
The suitable choice of the potential is a guarantee of the efficiency, accuracy, and fast convergence of the procedure.
For the hard sphere data, the natural choice is the hard sphere potential supplemented by the correction terms

\begin{equation}
\beta V(x)=(-V_a\, e^{-\kappa_a\,(x-1)}+V_r\,e^{-\kappa_r\,(x-1)})/x \  .
\label{11}
\end{equation}
($x$ here is $r/\sigma $).
Fitting four adjustable parameters $V_a$, $V_r$, $\kappa_a$ and $\kappa_r$ we estimate the corrections to the
hard sphere potential smaller than $\beta V(r) < 0.05$.
The calculated structure factor deviates from its ''experimental'' value less then $0.07 \%$!.

\begin{figure}
\begin{center}
\includegraphics[height=4in]{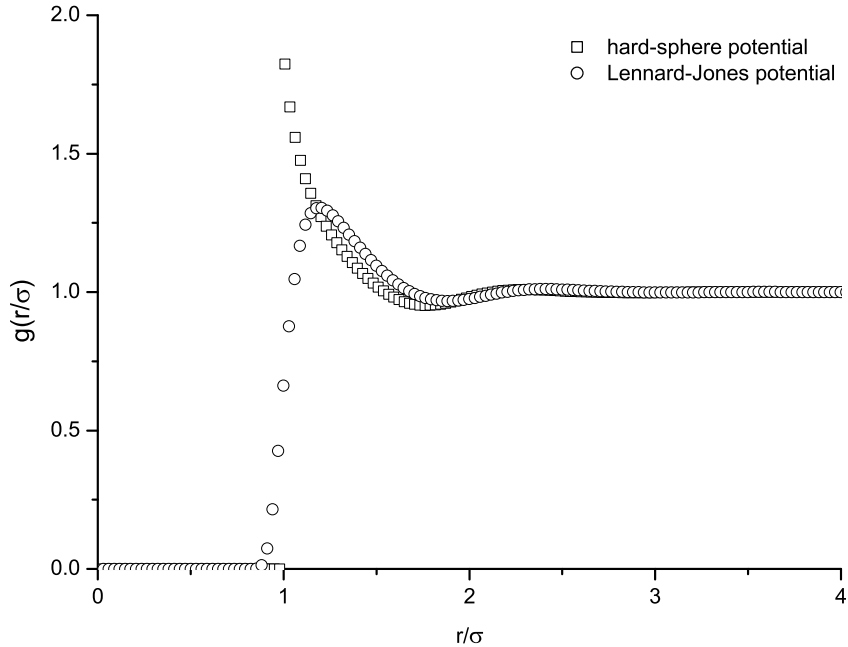}
\end{center}
\caption{Pair correlation function for hard-sphere (squares) and LJ (circles) potentials }
\label{f1}
\end{figure}

If we take the Lenard-Jones potential without the hard core part,
\begin{equation}
\beta V(r)=-V_a\, r^{-6}+V_r\, r^{-12} \ ,
\label{12}
\end{equation}
then the fitting to the ''experimental'' data (the exact OZ and PY equation solution for the hard spheres)
gives $V_a=0.4, \quad V_r=1.26$, and the deviation of the calculated structure factor from the ''experimental'' data is about
$0.7 \%$, i.e., 10 times worse then for the potential (\ref{11})).
A bit larger (but still not too bad) the differences between the both potentials take place if we
compare the computed and ''experimental'' pair correlation functions (see Fig. \ref{f1}.).
In own turn with the pair correlation function in hands, we can find such physically relevant quantity as the average
coordination number $N$

\begin{equation}
N=4\pi {\bar n} \int_{\sigma}^{1.2\,\sigma} dr\, r^2g(r)
\nonumber
\end{equation}
Here the integral is taken over the relatively narrow region of $r$ (around 
the main peak of correlation function at $r=\sigma $). For the potential (\ref{11}) $N \simeq 3.97$, whereas
for the Lenard-Jones potential $N \simeq 3.89$.
We conclude from these two pure methodical (but instructive) examples, that the both model effective potentials provide fairly good (although not ideal) data descriptions. The accuracy of the computed integral characteristics (like the average coordination
number $N$) is less impressive, it  is about
$2 \% $. If we were dealing with real experimental data (with finite systematic errors and noise), the accuracy
of our method can be considered as very satisfactory one.

\begin{figure}
\begin{center}
\includegraphics[height=4in]{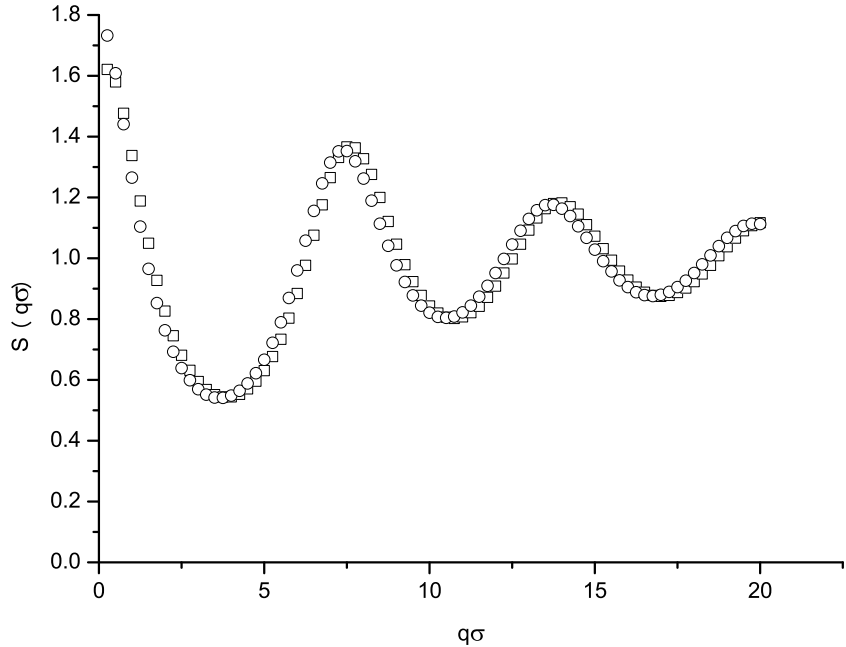}
\end{center}
\caption{Static structure factor for PMMA spheres with attraction caused by the depletion potential. Squares present the results obtained in \cite{Mur} 
and circles represent the results of the current approach}
\label{f2}
\end{figure}

Let us move now to the real experimental data. We take the scattering intensity data from the work \cite{Mur}
for the polymethylmetacrylate (PMMA) spheres. 
Using our methodology and depletion interaction potential induced by the polystyrene
globules dispersed in the solution \cite{Asakura}, we fitted all the experimental data presented in the work \cite{Mur}
with our new described above numerical procedure (which is simpler and faster than used in \cite{Mur}). 
One more advantage of the new approach, is that it is flexible and can be adapted for a rather wide range of the interaction potentials. Just as an illustration in the Fig. \ref{f2}, we plot the static structure factor calculated by the method of \cite{Mur} (black squares) and by our new method (red circles). Presented data correspond to volume fraction of PMMA particles
$\phi = 0.2$ and concentration of polyethylene glycol $c_p = 23 mg/L$. The radius of particles was fixed at the experimental value, and the interaction potential is the depletion potential. It is worth to note also that the standard method \cite{Mur}
gives the value for the potential amplitude $V_{a1} = -3.3 kT$, and the method of  our work leads to $V_{a2} = -3.55 kT$. 

\begin{figure}
\begin{center}
\includegraphics[height=4in]{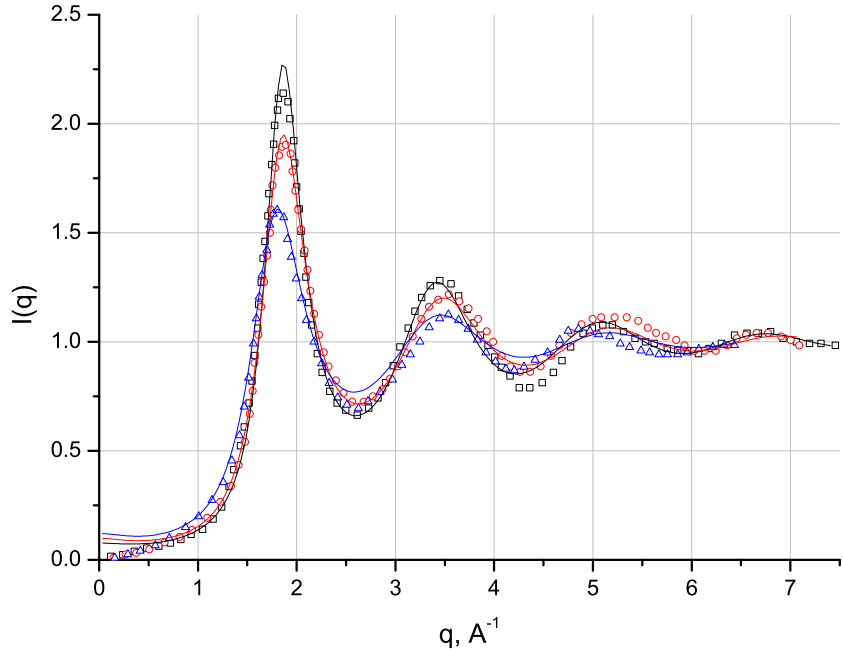}
\end{center}
\caption{Structure factor of liquid krypton at liquid-gas coexistence. The squares present the data for the temperature 133 K, circles - 153 K and triangles - 183 K. Solid lines present the result of fitting.}
\label{f3}
\end{figure}

Our next example is related to the neutron scattering data on liquid krypton \cite{Clayton}.
We reproduce the borrowed from the work \cite{Clayton} data on the scattering intensity in the Fig. \ref{f3}. 
The data cover the large interval of the scattering wave vectors $q\sigma \sim 1 \div 40$ (where $\sigma $
is estimated by the position of the main peak in the static structure factor).
Results of the fitting for the Lenard-Jones potential to the experimental 
data are presented in Fig. \ref{f3} by solid lines. The values of parameters determined from the same fitting 
are presented in Table 1. Four parameters were used as adjustable: scaling factor for the wave vector $\sigma $,
density $\bar n$ and the amplitudes of the Lenard-Jones potential $V_a$ and $V_r$.

\hspace{1 cm}Table 1
\begin{center}
\begin{tabular}{| c | c | c | c | c | c |}
   \hline
   $T $ & $\sigma $ & $\bar n$ & $V_a$ & $V_r$ & Coordination number \\ \hline
   133 & 3.91 & 0.72 & 2.56 & 1.91 & 7.3 \\ \hline
   153 & 3.55 & 0.68 & 3.11 & 3.77 & 6.5 \\ \hline
   183 & 3.55 & 0.68 & 2.64 & 3.6 & 4.7  \\ \hline
\end{tabular}
\end{center}

By physical arguments (or even common wisdom) for the krypton one should expect the Lenard-Jones interaction
potential provides an adequate description of the system. Surprisingly enough (see the Fig. \ref{f3}) it is not the case (as it was mentioned
in \cite{Clayton} and as we have confirmed by our own computation). Something is evidently wrong.
In our opinion, the catch is in a small $q$ region. As it is well known (see e.g., \cite{HM06}) there is 
the exact thermodynamic relation
\begin{equation}
S(0)=k_B T \left(\frac {\partial {\bar n}}{\partial P}\right) _T \ .
\label{14}
\end{equation}
We take the values entering (\ref{14}) parameters from \cite{Clayton} and the handbook \cite{hand} and calculate the structure factor at $q=0$
for a few temperature points along the liquid-gas coexistence line:
$S(q=0, T=133 K)\approx 0.076$, $S(q=0, T=153 K)\approx 0.13$, $S(q=0, T=183 K)\approx 0.459$.

\begin{figure}
\begin{center}
\includegraphics[height=4in]{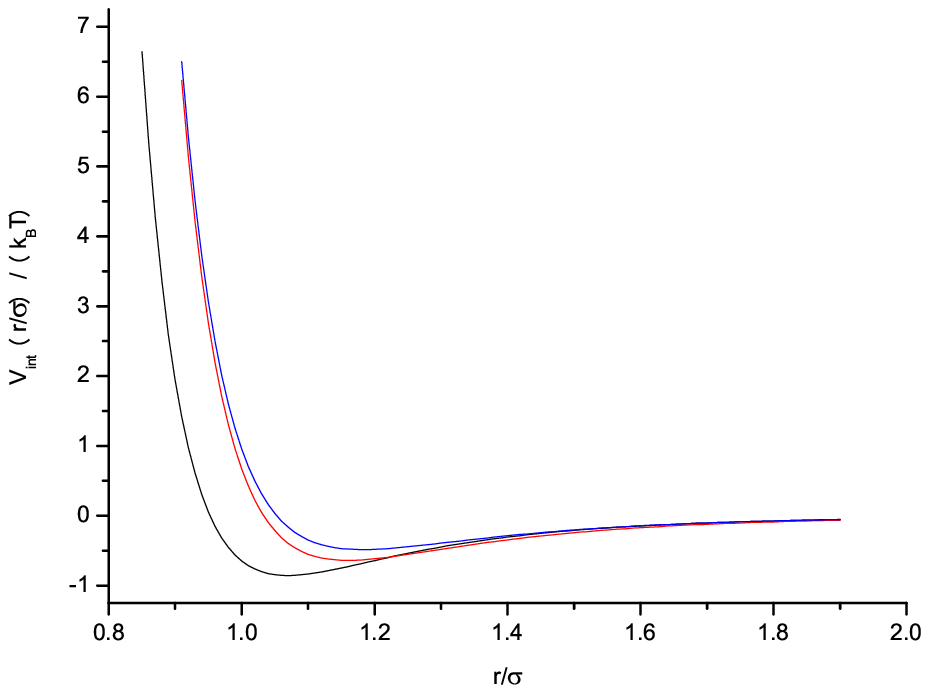}
\end{center}
\caption{Lenard-Jones interaction potential for liquid krypton. Black line correspond to
133 K, red line - 153 K and blue line - 183 K.}
\label{f4}
\end{figure}

With these so-to-speak exact values for the $S(q=0)$, one has to perform the fitting of the scattering data in the broad range of the wave vectors.
We present the results of our approach model fitting with the Lenard-Jones inter-particle potential.
The magnitudes of all needed and computed parameters are presented in the table 1.
To find all three correlation functions we utilize the PY closure relation, but similar by quality fitting 
can be obtained also with the HNC closure relation.
As we said already the separate important task is to find (estimate) the interaction potential.
We plot the result obtained by our method in the Fig. \ref{f4}.

\begin{figure}
\begin{center}
\includegraphics[height=4in]{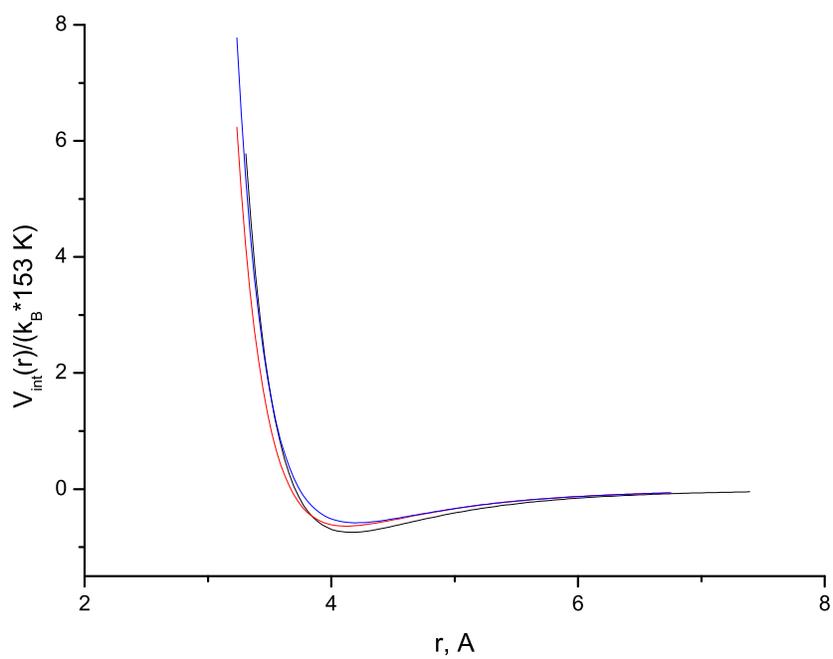}
\end{center}
\caption{Interaction potential for liquid krypton, rescaled to the temperature 153 K. Lines are the same as in Fig. \ref{f4}.}
\label{f5}
\end{figure}

The curves in the Fig. \ref{f4} present the potential divided by $k_BT$. All three curves can be rescaled and collapsed into a single universal (master) curve. The result of such
procedure is presented in Fig. \ref{f5}.

\section{Conclusion and Perspectives}
\label{con}

In recent years there has been an upsurge in interest in structural investigations of various
colloidal suspensions and molecular liquids (see e.g., \cite{LN11}, \cite{last} and references therein), although the problem itself is anything but new.
The van der Waals theory is the cornerstone of the current understanding of fluid structures and phase behavior.
Although modern experimental and numeric methods of structure investigations can provide very detailed information about
many features and properties of liquid state, no simple way to relate the both, because the data depend on a number of 
material and microscopic parameters only barely known. To get such insight one has to rely on the traditional methods of the statistical physics, namely, having in mind the fluid state structure, it is necessary to solve the exact OZ equation. Unfortunately, here we face to a problem, because the equation is not in a closed
form: it contains two unknown functions. In a general case (molecular liquids, or not very dilute colloidal dispersions)
it is not possible to derive by a regular theoretical method the needed closure relations. The most popular closure
relations (PY and HNC) are basically a sort of self-consistent extrapolations from the very dilute dispersion limit.
The PY closure allows to find the exact analytical solution of the OZ equation for the dispersion of hard spheres. Moreover,
the PY closure leads to a reasonably good description of the experimental data for the dispersions somehow close to the hard
sphere ones. Then it is tempting to try to describe experimental data for a more broad class of dispersions using the trial PY
or HNC ansatz for the correlation functions and computing perturbatively corrections to the ansatz to fit the data.
Unfortunately such a perturbative approach is not always work, and besides leads to not very efficient and fast
computations. To find how to overcome the difficulty, there is no way to improve the theory all the more that it is impossible
for a system without any small parameter.
As one can expect this intermediate range of parameters 
is the most difficult one to treat theoretically, all the more, analytically. 

Instead, we propose a simple and working instrument, combining the theory solution
to the OZ equation for the function $\gamma (r)$ and experimentally measured scattering intensity.
It is based on the observation that it is
possible to determine the smooth function $\gamma (r) \equiv h(r) - c(r)$ directly from the experimental data for the structure factor
(scattering intensity data for the system under consideration and from its diluted state). Then our  method enables to compute the correlation functions and the inter-particle potential by using the OZ equation with a closure relation. The procedure is robust and simple, and does not use any ansatz for the correlation functions.

The delicate issue in the procedure is the choice of the functional form for the interaction potential.
The accuracy of the resulting description (especially as it concerns to the integral characteristics) depends
essentially on the choice.
We illustrate on some model and real experimental examples, how the method works. 
For relatively small number of the fitting parameters, namely the parameters entering the interaction potential
and the scaling factor $\sigma $ for the wave vectors, the results are very satisfactory.
In this sense our approach and the model can be considered as the minimal model, which is just at the border
between those that are too primitive to fit even qualitatively
the data, and those that fit the data too well by using too many parameters.

\vspace{0.5cm}

Acknowledgements

\vspace{0.5cm}

We acknowledge stimulating discussions with T.Narayanan, inspired this work.
A.M. thanks RFFI grant number 15-08-07727
for partial financial support.

\end{document}